\documentclass[aps,showpacs,nofootinbib]{revtex4}
\usepackage{epsf,latexsym}
\usepackage{amsfonts}

\begin{document}

\title{Looking at spacetime atoms from within the Lorentz sector}
\author{Alessandro Pesci\footnotetext{e-mail: pesci@bo.infn.it}}
\affiliation
{INFN-Bologna, Via Irnerio 46, I-40126 Bologna, Italy}

\begin{abstract}
Recently, a proposal has been made to figure out the
expected discrete nature of spacetime at the smallest scales
in terms of atoms of spacetime,
capturing their effects
through a scalar $\rho$, function of the point $P$
and the vector $v^a$ at $P$,
expressing
their density.
This has been done
in the Euclideanized space
one obtains through analytic continuation
from Lorentzian sector at $P$.
$\rho$ is defined in terms of a peculiar `effective' metric $q_{ab}$,
also recently introduced,
which stems from a careful request that $q_{ab}$ coincides
with $g_{ab}$ at large (space/time) distances, but gives finite distance
in the coincidence limit.

This work reports on
an attempt to introduce a definition
of $\rho$ directly in the Lorentz sector.
This turns out to be not a so trivial task,
essentially because of the null case, i.e. when $v^a$ is null,
as in this case we lack even a concept of $q_{ab}$.
A notion for $q_{ab}$ in the null case is here proposed
and an expression for it is derived.
In terms of it,
an expression for $\rho$ can be derived,
which turns out to coincide with what obtained
from analytic continuation.
This, joined with the consideration of timelike/spacelike cases,
potentially completes a description of $q_{ab}$ and $\rho$
within Lorentz spacetimes.  
\end{abstract}


\maketitle

$ $
\section{Staying in the Lorentz sector}

Previous works \cite{KotE, KotF, StaA}
have proved it quite useful to introduce 
a peculiar sort of effective metric, also called qmetric,
which acts to some extent as a metric, at the same time implementing
the existence of a finite limiting distance $L$
between two events in their coincidence limit,
this way implementing intrinsic discreteness of spacetime
and still not abandoning the benefits, for calculus, associated
to a continuous description of spacetime.
One result one gets this way is
the possibility to analitically provide a notion 
of degrees of freedom or of number of (quantum) states
of spacetime \cite{Pad04, Pad06, Pad08, Pad10},
fact which paves the way to a statistical description
of field equations, and then to express the basic tenets of gravity
using as proper language thermodynamics (as opposed to geometry)
\cite{Pad04}.

Key to the notion of degrees of freedom or of number of states
of spacetime is a quantity, denoted here $\rho$,
defined in terms of $(D-1)$-dimensional areas
(spacetime is assumed $D$-dimensional) of hypersurfaces
formed by points at assigned
distance from some point $P$
in the space coming from
Euclideanisation of original spacetime around $P$.
The fundamental feature here
is that, according to the effective metric,
these $(D-1)$-areas remain finite
in the coincidence limit in which the hypersurfaces shrink to $P$
\cite{Pad04} (and clearly, one would expect some analogous results
do hold true in Lorentzian sector). 

While this of Euclideanisation
might be a point of merit,
providing insight perhaps into what the structure of the
metric might be at the smallest scales, 
the results one thus obtains have anyway
to be retranslated back
to Lorentz sector, since this is what we
actually deal with in natural phenomena at ordinary scales.
The aim of present study is to develop
a concept of $\rho$ in the Lorentz sector directly,
i.e. with no reliance on Euclideanised space.

A partial result in this direction
has been already presented in \cite{PesK}.
There,
a notion of $\rho$ for timelike geodesics has been introduced
and its expression has been derived.
What is left out is the most part of the work to do,
namely the consideration of null geodesics
(the discussion of spacelike case is almost identical
to the timelike one).
This of null geodesics, is the case we try to study here.

\section{$\rho$ for timelike/spacelike geodesics}

Let us start by recalling
what we can do with timelike/spacelike geodesics.
We briefly rephrase what is reported in \cite{PesK}
for timelike case, using here a notation which encompasses
both the timelike and the spacelike case at one stroke. 
We consider timelike/spacelike geodesics
through a generic point $P$ in spacetime,
and introduce the two hypersurfaces
$\Sigma_\epsilon(P, l)$,
$\epsilon = +1$ for spacelike geodesics
and $\epsilon = -1$ for timelike ones,
of all points $p$
at assigned squared distance from $P$:

\begin{eqnarray}
  \Sigma_\epsilon(P, l) = \big\{p: \ \epsilon\sigma^2(p, P) = l^2 \big\},
  \nonumber
\end{eqnarray}
where 
$
\sigma^2(p, P)
$
is the squared geodesic distance between $P$ and $p$
($\sigma^2(p, P) = 2 \Omega(p, P)$,
with $\Omega(p, P)$ the Synge
world function \cite{Syn}),
and $l = \sqrt{l^2}$ non-negative.

Proceeding analogously
to the Euclidean definition,
$\rho$
is given in terms of generic/flat ratio of element of areas
on $\Sigma_\epsilon(P, l)$,
as measured according to the effective metric,
in the limit $l \rightarrow 0$.
For each assigned normalised vector $n^a$ at $P$
($n^a n_a = \epsilon$),
we consider the intersection point $p$ between
the geodesic $\mu(n^a)$
with tangent at $P$ $t^a(P) = n^a$
and the hypersurface $\Sigma_\epsilon(P, l)$.
Calling $y^i$, $i = 1, ..., D-1$
coordinates on $\Sigma_\epsilon(P, l)$ 
such that $y^i(p) =0$,
we consider a segment $I$ of hypersurface $\Sigma_\epsilon(P, l)$
around $p$, defined as
$
I = \{dy^i\},
$
where $dy^i$ are thought as fixed when $l$ is varied.
The $(D-1)$-dimensional area of $I$ is

\begin{eqnarray}
  d^{D-1}V(p) = \sqrt{- \epsilon h(p)} \ d^{D-1}y, \nonumber
\end{eqnarray}
where $h_{ij}$ are the components of the metric
on $\Sigma_\epsilon(P, l)$
in the coordinates $y^i$,
metric which coincides with that induced
by spacetime metric $g_{ab}$.
What we have to consider is the area $[d^{D-1}V]_q$ of $I$ 
as measured through the effective metric $q_{ab}$.

The effective metric
is described \cite{KotE, StaA}  in terms of the bitensor
$q_{ab}(p, P)$
which stems from requiring the squared
geodesic distance
$\sigma^2$
gets modified into
$
\sigma^2 \rightarrow [\sigma^2]_q =
      {S_L}(\sigma^2) 
$
with
{\bf (R1)} $S_0 = \sigma^2$,
{\bf (R2)} ${S_L}(0^\pm) = \pm L^2$,
and
{\bf (R3)}
the kernel $G(\sigma^2)$ of the d'Alembertian
gets modified into
$G(\sigma^2) \rightarrow [G]_q(\sigma^2) = G({S_L})$
in all maximally symmetric spacetimes.
These requirements give, for spacelike or timelike geodesics,
the expression

\begin{eqnarray}\label{qab}
  q_{ab}(p, P) = A(\sigma^2) g_{ab}(p) +
  \epsilon \Big(\frac{1}{\alpha(\sigma^2)} - A(\sigma^2)\Big) t_a(p) t_b(p),
\end{eqnarray}
where $t^a$ is the normalized tangent vector
($g_{ab} t^a t^b = \epsilon$),
not going to change its timelike or spacelike character
when in the qmetric,

\begin{eqnarray}\label{A}
  A = \frac{S_L}{\sigma^2}
  \Big(\frac{\Delta}{\Delta_S}\Big)^\frac{2}{D-1},
\end{eqnarray}

\begin{eqnarray}\label{alpha}
  \alpha = \frac{S_L}{\sigma^2 (S^\prime_L)^2}
\end{eqnarray}
($^\prime$ indicates differentiation with respect
to the argument $\sigma^2$),
where
\begin{eqnarray}\label{vanVleck}
  \Delta(p, P) = - \frac{1}{\sqrt{g(p) g(P)}}
  {\rm det}\Big[-\nabla^{(p)}_a \nabla^{(P)}_b \frac{1}{2} \sigma^2(p, P)\Big]
\end{eqnarray}
is the van Vleck determinant
(\cite{vVl, Mor, DeWA, DeWB}; see \cite{Xen, VisA, PPV})
which is a biscalar,
and the biscalar $\Delta_S(p, P)$ is
$\Delta_S(p, P) = \Delta({\tilde p}, P)$,
where $\tilde p$ is
that point on the geodesic through $P$ and $p$
(on the same side of $p$ with respect to $P$)
which has $\sigma^2({\tilde p}, P) = S_L(p, P)$.
$\alpha$ is determined by the request
that the formula for squared geodesic distance

\begin{eqnarray}\label{HJ}
  g^{ab} \partial_a\sigma^2 \partial_b\sigma^2 = 4 \sigma^2
\end{eqnarray}
(Hamilton-Jacobi equation)
gets transformed into
$q^{ab} \partial_a S_L \partial_b S_L = 4 S_L$;
$A$ by the request {\bf R3}.

From
the effective metric $[h_{ab}]_q(p, P)$ induced by $q_{ab}(p, P)$
at $p$ on $\Sigma_\epsilon(P, l)$,
we get 
the effective-metric $(D-1)$-dimensional area of $I$ as
\begin{eqnarray}
 [d^{D-1}V]_q(p, P)
 = \Big[\sqrt{- \epsilon h}\Big]_q(p, P) \ d^{D-1}y . \nonumber
\end{eqnarray}
As in the Euclidean approach,
$\rho$ can then be defined as the ratio
of effective-metric $(D-1)$-dimensional area of $I$
for the actual metric configuration, $[d^{D-1}V]_{q(g)}(p, P)$, 
to what we would have were spacetime flat, $[d^{D-1}V]_{q(\eta)}(p, P)$
($\eta_{ab}$ is Minkowski metric), in the
limit $p \rightarrow P$ along $\mu(n^a)$,
i.e.

\begin{eqnarray}\label{rho_def}
  \rho(P, n^a) =
  \bigg(\lim_{p \rightarrow P}
  \frac{[d^{D-1}V]_{q(g)}(p, P)}{[d^{D-1}V]_{q(\eta)}(p, P)}\bigg)
  _{\mu(n^a)} .
\end{eqnarray}

$\rho$ is then derived
in terms of the quantities $A$ and $\alpha$ defining the effective metric.
The effective metric $[h_{ab}]_q$ induced by $q_{ab}$
turns out to be

\begin{eqnarray}
  [h_{ab}]_q(p, P) = A(\sigma^2) h_{ab}(p)
  \nonumber
\end{eqnarray}  
\cite{KotG},
which implies

\begin{eqnarray}
  \Big[\sqrt{- \epsilon h}\Big]_q(p, P) = A(\sigma^2)^{\frac{D-1}{2}}
  \sqrt{- \epsilon h(p)},
  \nonumber
\end{eqnarray}  
and then

\begin{eqnarray}
  [d^{D-1}V]_q(p, P)
  &=& A(\sigma^2)^{\frac{D-1}{2}} d^{D-1}V(p), \nonumber
\end{eqnarray}
where $d^{D-1}V(p)$ indicates the proper area of $I$
according to the ordinary metric.
Here we see that
only $A$, and not $\alpha$,
is actually involved
in the determination of $\rho$.

Introducing
on $\Sigma_\epsilon$, in a neighbourhood of $p$,
mutually orthogonal coordinates $z^i$ 
such that,
chosen any one of them, $z^{\bar i}$,
it can be written in the form
$z^{\bar i} = l \eta$
with the parameter $\eta$ such that $l d\eta$ is proper distance
or proper-time difference,
and chosing as $I$ the (hyper)cube ${dz^i}$
defined by $dz^i = l d\eta, \forall i$,
we obtain

\begin{eqnarray}
  [d^{D-1}V]_q(p, P)
  &=& A(\sigma^2)^{\frac{D-1}{2}} l^{D-1}
      \big(1 + {\cal O}(l^2)\big) (d\eta)^{D-1}
      \nonumber
\end{eqnarray}
where the ${\cal O}(l^2)$ term
represents the effects of curvature
(and is thus of course absent in flat case),
and clearly
$l = \sqrt{\epsilon \sigma^2}.$
Using the expression (\ref{A}) for $A$,
we get

\begin{eqnarray}
  [d^{D-1}V]_q(p, P)
  = [\epsilon S_L]^{\frac{D-1}{2}}
      \frac{\Delta(p, P)}{\Delta_S(p, P)} \
      \big(1 + {\cal O}(l^2)\big) (d\eta)^{D-1}   \nonumber
\end{eqnarray}
and,
in the limit $p \rightarrow P$ along $\mu(n^a)$,

\begin{eqnarray}
  \lim_{l \rightarrow 0}
  \ [d^{D-1}V]_q(p, P)
  = L^{D-1} \frac{1}{\Delta_L(P, n^a)} (d\eta)^{D-1},     \nonumber
\end{eqnarray}  
with
$\Delta_L(P, n^a) = \Delta({\bar p}, P)$, where $\bar p$ 
is that point on geodesic $\mu(n^a)$ (on the side in the direction $n^a$)
which has $l = L$.

This shows that both the numerator and the denominator
in expression (\ref{rho_def}) remain non vanishing
in the coincidence limit $p \rightarrow P$,
exactly as it happens in Euclidean case.
Since for flat spacetime $\Delta = 1$ identically and then also
$\Delta_L = 1$,
we have finally

\begin{eqnarray}\label{rho}
  \rho(P, n^a)
  = \frac{1}{\Delta_L(P, n^a)},
\end{eqnarray}
where the $\Delta_L$ is that of generic metric $g_{ab}$. 
The scope of this exact expression for $\rho$
clearly includes strictly Riemannian manifolds (as that from Euclideanisation). 

Expanding $\Delta(p, P)$ in powers of $l$
(\cite{DeWA}; \cite{Xen, VisA, PPV}),

\begin{eqnarray}\label{expansion}
  \Delta(p, P) = 1 + \frac{1}{6} l^2 R_{ab} t^a t^b
  + o\big(l^2 R_{ab} t^a t^b\big),
\end{eqnarray}  
($t^a t_a =\epsilon)$ 
gives

\begin{eqnarray}\label{DeltaL}
 \Delta_L(P, n^a) =  
 1 + \frac{1}{6} L^2 R_{ab}(P) n^a n^b + o\big(L^2 R_{ab}(P) n^a n^b\big),
\end{eqnarray}
and 

\begin{eqnarray}\label{rho_expansion}
  \rho(P, n^a) =
  1 -\frac{1}{6} L^2 R_{ab}(P) n^a n^b + o\big(L^2 R_{ab}(P) n^a n^b\big).
\end{eqnarray}
Again, this identically applies also to Riemannian manifolds
(as that from Euclideanisation),
and its form coincides with the expansion obtained
\cite{Pad04, Pad06, Pad08, Pad10} defining $\rho$
in the Euclideanised space.

\section{qmetric and null geodesics}

If we try to extend the scope of effective metric approach
to include null geodesics,
we have that expression (\ref{qab}) becomes ill defined in this case
since
$\sigma^2 =0$ all along any null geodesic,
and in principle we are then in trouble.
We notice however the following.
Any affine parametrization $\lambda$ of a null geodesic can be thought of
as a measure of distance along the geodesic performed
by a canonical observer
picked up at a certain point $x$ of the geodesic and parallel transported
along the geodesic.
Since, when going to the effective metric $q_{ab}$,
the squared distance in the coincidence limit is the finite value
$\epsilon L^2$ (request {\bf R2} above),
we could expect the effect of the qmetric in the null case is
to induce a mapping of the parametrization $\lambda$ to a new parametrization
$\tilde \lambda = \tilde \lambda(\lambda)$,
with ${\tilde \lambda} \rightarrow L$ when
$\lambda(p, P) \rightarrow 0$, i.e. when $p \rightarrow P$.
In analogy with the spacelike/timelike case,
we can then think to give
an expression for $q_{ab}(p, P)$ when $p$ is on a null
geodesic from $P$
in terms of two functions $\alpha_\gamma = \alpha_\gamma(\lambda)$
and $A_\gamma = A_\gamma(\lambda)$ defined on the geodesic,
and determined by a condition on the squared geodetic distance
and on the d'Alembertian.
In other words,
this suggests we assume
that the effects of the existence of a limiting length  
are captured by an effective metric bitensor $q_{ab}$ as above,
with its expression on a null geodesic stemming from requiring
the affine parametrization $\lambda$ gets modified
into $\lambda \rightarrow [\lambda]_q = {\tilde \lambda}(\lambda)$
with
{\bf (G1)}
${\tilde \lambda} = \lambda$ if $L = 0$
(or ${\tilde \lambda} \simeq \lambda$ when $\lambda \rightarrow \infty$),
{\bf (G2)}
${\tilde \lambda}(0^+) = L$,
and
{\bf (G3)}
the kernel $G(\sigma^2)$ gets modified into $[G]_q(\sigma^2) = G(S_L)$
in all maximally symmetric spacetimes, i.e {\bf (G3)} coincides
with {\bf (R3)} above on null geodesics.

We see that dealing with the null case appears quite not
so obvious,
in that we are forced to rewrite for this case from scratch
the rules to go to the qmetric given a metric, in terms of an
affine parameter $\lambda$ defined on null geodesics only,
i.e. $q_{ab}$ is defined strictly on null geodesics
and knows nothing outside them.
And this, morover,
leads to the tricky circumstances
that the operators we look at when constraining the expression
for $q_{ab}$ (e.g. the d'Alembertian) should be considered in a form
which does not hinge on any knowledge,
regarding the elements which enter the definition of the operator itself
(directional derivatives, vectors),
of what happens outside the $(D-1)$-dimensional submanifold
swept by all the null geodesics emanating from a point.

Let $\gamma$ be a null geodesic through $P$,
with affine parameter
$\lambda = \lambda(p, P)$ with $\lambda(P, P) = 0$,
and null tangent vector $l^a = \frac{dx^a}{d\lambda}$,
i.e. $\nabla_a (\sigma^2) = 2 \lambda l_a$ (see e.g. \cite{PPV}). 
We introduce a canonical observer at $P$, with velocity $V^a$,
such that $l_a V^a = -1$. By parallel transport of the observer
along $\gamma$, this relation extends all along $\gamma$.
We affinely parametrize any other null geodesic $\hat \gamma$
which goes through $P$, and require ${\hat l}_a V^a = -1$.
What we obtain this way,
is
a $(D-1)$-dimensional congruence $\Gamma$ of null geodesics
emanating from $P$ which is affinely parametrized
and has deviation vectors orthogonal to the geodesics.
We introduce a second null vector $m^a$ at $P$, defined by
$m^a \equiv 2 V^a - l^a$, and parallel transport it along the geodesic.
This gives $m_a V^a = -1$ and $m_a l^a = -2$ all along $\gamma$.
The vector $m^a$ does depend on the observer we have chosen.

Let $q_{ab}(p, P)$, $p$ on $\gamma$, be of the form

\begin{eqnarray}\label{qab_null}
  q_{ab} = A_\gamma g_{ab} -\frac{1}{2}
          \Big(\frac{1}{\alpha_\gamma} - A_\gamma\Big) (l_a m_b + m_a l_b).
\end{eqnarray}
From $q^{ab} q_{bc} = \delta^a_c$, we get

\begin{eqnarray}
  q^{ab} = \frac{1}{A_\gamma} g^{ab} + \frac{1}{2}
          \Big(\frac{1}{A_\gamma} - \alpha_\gamma\Big) (l^a m^b + m^a l^b),
\end{eqnarray}
where $l^a = g^{ab} l_b$, $m^a = g^{ab} m_b$. 
Notice that $q^{ab} l_a l_b = 0$, and the geodesic is null also according
to the qmetric.

Our first task is to determine the form of $\alpha_\gamma$
from the condition on squared distance.
To this aim,
the direct use, on $\gamma$, of (\ref{HJ}) is of no help,
since $\sigma^2$ is identically vanishing there.
We notice, however,
that the condition
on squared distance can be reformulated in an alternative way.
Let us consider a spacelike geodesic.
We can write

\begin{eqnarray}\label{sigma}
  \sigma(p, P) = \int_0^{\chi(p, P)} \sqrt{g_{ab} v^a v^b} d\chi,
\end{eqnarray}
being the integral on the geodesic,
with $\sigma$ the distance,
$\chi$ parameter which is not necessarily affine,
and $v^a = \frac{dx^a}{d\chi}$. 
This gives

\begin{eqnarray}
  \nonumber
  \sigma(p, P)
  = \int_0^{\chi(p, P)} \sqrt{g_{ab}
                 \frac{dx^a}{d\chi} \frac{dx^b}{d\chi}} d\chi
  = \int_0^{\sigma(p, P)} d\sigma
  = \int_0^{\sigma(p, P)} \sqrt{g_{ab}
                 t^a t^b} d\sigma,   
\end{eqnarray}
with $t^a$ tangent to the geodesic with $t_a t^b = 1$.
What this means is that total proper distance can be given as the sum of
elementary proper distances associated to the differences
$d\chi$ or $d\sigma$ of the parameter. 
Going to the qmetric,
one would then require

\begin{eqnarray}
  \sqrt{S_L} = \int_0^{\sigma(p, P)} \sqrt{q_{ab} t^a t^b} d\sigma.
\end{eqnarray}
From
$ q_{ab} t^a t^b = \frac{1}{\alpha}$
(eq. (\ref{qab})), this gives

\begin{eqnarray}
  \nonumber
  \sqrt{S_L} = \int_0^{\sigma(p, P)} \frac{1}{\sqrt{\alpha}} d\sigma,
\end{eqnarray}
from which

\begin{eqnarray}
  \nonumber
  \frac{d\sqrt{S_L}}{d\sigma} = \frac{1}{\sqrt{\alpha}},
\end{eqnarray}
and then

\begin{eqnarray}
  \nonumber
  \frac{dS_L}{d(\sigma^2)} \frac{\sigma}{\sqrt{S_L}}
  = \frac{1}{\sqrt{\alpha}},
\end{eqnarray}
which gives for $\alpha$ the expression (\ref{alpha}).

In the case of null geodesic,
the affine parameter $\lambda$
can be written,
looking at (\ref{sigma}),
as

\begin{eqnarray}
  \lambda(p, P) = \int_0^{\lambda(p, P)} \sqrt{g_{ab} X^a X^b} d\lambda
\end{eqnarray}
where 
$X^a \equiv l^a - V^a$ is unit spacelike, in the direction
of propagation of the light ray according to the canonical observer $V^a$,
and the integral is on the null geodesic.
When going to the qmetric,
what we should require is then

\begin{eqnarray}
  {\tilde \lambda}(p, P) = \int_0^{\lambda(p, P)}
                         \sqrt{q_{ab} X^a X^b} d\lambda.
\end{eqnarray}  
From the expression (\ref{qab_null})
for $q_{ab}$ on null geodesics,
we get
$
q_{ab} X^a X^b = \frac{1}{\alpha_\gamma},
$  
and

\begin{eqnarray}
 \nonumber
 {\tilde \lambda}(p, P) =  \int_0^\lambda(p, P) \frac{1}{\alpha_\gamma} d\lambda.
\end{eqnarray}  
This gives

\begin{eqnarray}
  \nonumber
  \frac{d{\tilde \lambda}}{d\lambda} = \frac{1}{\sqrt{\alpha_\gamma}},
\end{eqnarray}
and then

\begin{eqnarray}\label{alpha_gamma}
  \alpha_\gamma = \frac{1}{(d{\tilde \lambda}/d\lambda)^2}. 
\end{eqnarray}  

As for the determination of $A_\gamma$,
we have to refer to {\bf G3},
i.e we consider the d'Alembertian
in maximally symmetric spaces on null geodesics.
What we try first, is to find out some convenient expression for the
d'Alembertian.
Due to maximal symmetry,
we can think in terms of $f = f(\sigma^2)$
and write

\begin{eqnarray}
  \Box f &=& \nabla_a \nabla^a f
  \nonumber \\
  &=& \nabla_a\Big(\partial^a\sigma^2 \frac{df}{d\sigma^2}\Big)
  \nonumber \\
  &=& \big(\nabla_a  \partial^a\sigma^2\big) \frac{df}{d\sigma^2} +
  \big(\partial^a\sigma^2\big) \partial_a \frac{df}{d\sigma^2}
  \nonumber \\
  &=& \big(\nabla_a  \partial^a\sigma^2\big) \frac{df}{d\sigma^2} +
  \big(\partial^a\sigma^2\big)
  \big(\partial_a\sigma^2\big) \frac{d^2f}{d(\sigma^2)^2}.
  \nonumber
\end{eqnarray}
When going to null geodesic $\gamma$,
$ \big(\partial^a\sigma^2\big) \big(\partial_a\sigma^2\big)
\rightarrow (2 \lambda l^a) (2 \lambda l_a) = 0 $
and we get

\begin{eqnarray}
  \nonumber
  \Box f = \big(\nabla_a  \partial^a\sigma^2\big) \frac{df}{d\sigma^2}.
\end{eqnarray}
At a point ${\tilde p}$ close to $\Gamma$
but, possibly, not exactly on it,
we can write (cf. \cite{VisA})

\begin{eqnarray}
  \nonumber
  \partial^a\sigma^2({\tilde p}) =
    2 \lambda \ l^a_{|{\tilde p}} + 2 \nu \ m^a_{|\tilde p}, 
\end{eqnarray}
where $\lambda$ and $\nu$ are curvilinear null coordinates of $\tilde p$
(there is a unique point $p$ on $\Gamma$ from which $\tilde p$
is reachable
through a null geodesic $\gamma'$ with tangent $m^a$ at $p$;
$\nu$ is the affine parameter of ${\tilde p}$ along $\gamma'$,
with $\nu(p) = 0$),
$l^a_{|{\tilde p}}$ and $m^a_{|\tilde p}$ are $l^a$ and $m^a$ parallel
transported along $\gamma'$ from $p$ to $\tilde p$.
This gives, on $\gamma$,

\begin{eqnarray}\label{nabla_partial}
  \nabla_a\partial^a\sigma^2
  &=& 2 \ \big(\lambda \nabla_a l^a
  + l^a \partial_a\lambda + m^a \partial_a\nu\big)
  \nonumber \\
  &=& 2 \ \big(\lambda \nabla_a l^a + 2),
\end{eqnarray}
and then

\begin{eqnarray}
  \Box f
  =
  \big(4 + 2 \lambda \nabla_a l^a\big) \frac{df}{d\sigma^2}
  =
  \big(4 + 2 \lambda \nabla_i l^i\big) \frac{df}{d\sigma^2}, 
\end{eqnarray}
$i = 1, ..., D-1$ indices of components on $\Gamma$.
Here, we emphasized the fact that,
since the covariant derivative of $l^a$ along $\gamma'$ is 0,
$\nabla_a l^a$
is completely defined within $\Gamma$
and coincides with the expansion of $\Gamma$,
$\nabla_a l^a = \nabla_i l^i$.

Going to the qmetric,
the geodesic $\gamma$ remains null,
and we have

\begin{eqnarray}\label{Box_q1}
  [\Box]_q f &=& \big(4 + 2 [\lambda \nabla_a l^a]_q\big) \frac{df}{d\sigma^2}
  \nonumber \\
  &=& \big(4 + 2 [\lambda]_q \ [\nabla_a l^a]_q\big) \frac{df}{d\sigma^2}
  \nonumber \\
  &=& \big(4 + 2 {\tilde \lambda}
  \ [\nabla_a l^a]_q\big) \frac{df}{d\sigma^2}.
\end{eqnarray}
Here
$
[l^a]_q = dx^a/d{\tilde \lambda} = (d\lambda/d{\tilde \lambda}) \ l^a.
$
As for the divergence,
we have
$
[\nabla_a l^a]_q =
[(\partial_a + {\Gamma^b}_{ab}) l^a]_q,
$
with,
from
$
{\Gamma^a}_{bc} = \frac{1}{2} g^{ad}
(-\partial_d g_{bc} + \partial_c g_{bd} +\partial_b g_{dc}),
$

\begin{eqnarray}
  [{\Gamma^a}_{bc}]_q
  &=& \frac{1}{2} q^{ad}
  (-\partial_d q_{bc} + \partial_c q_{bd} +\partial_b q_{dc})
  \nonumber \\
  &=& \frac{1}{2} q^{ad}
  (-\nabla_d q_{bc} + 2 \nabla_{\left(b\right.} q_{\left.c\right)d}) + {\Gamma^a}_{bc}
  \nonumber
\end{eqnarray}
(cf. \cite{KotG}),
provided we can give meaning to the differentiations involved.
A potential problem we have in these expressions, in fact,
is that the differentiations which are involved bring us
to leave $\Gamma$;
we have to remember however that we have further to contract with $[l^a]_q$,
and, on contracting,
the differentiations which turn out to be really present in the q-divergence
can actually result {\it on} $\Gamma$,
as we would expect from the fact that the qmetric does respect the null nature
of $\Gamma$, with $[l^a]_q$ being tangent to the generators of it. 
Last formula gives

\begin{eqnarray}
  [{\Gamma^b}_{ab}]_q
  &=&
  \frac{1}{2} q^{bc}
  (-\nabla_c q_{ab} + 2 \nabla_{\left(a\right.} q_{\left.b\right)c}) + {\Gamma^b}_{ab}
  \nonumber \\
  &=&
  \frac{1}{2} q^{bc} \nabla_a q_{bc} + {\Gamma^b}_{ab},
  \nonumber
\end{eqnarray}
and

\begin{eqnarray}
  \nonumber
  [\nabla_a l^a]_q =
  \nabla_a\Big(\frac{d\lambda}{d{\tilde \lambda}} l^a\Big)
  + \frac{1}{2} q^{bc} (\nabla_a q_{bc})
  \ \frac{d\lambda}{d{\tilde\lambda}} \ l^a.
\end{eqnarray}
This espression openly shows that
all differentials are indeed taken on $\Gamma$. 
Using formula (\ref{qab_null}) for $q_{ab}$,
direct computation gives

\begin{eqnarray}
  [\nabla_a l^a]_q
  &=& \frac{d\lambda}{d{\tilde\lambda}} \ \nabla_i l^i +
  \frac{d}{d\lambda} \Big(\frac{d\lambda}{d{\tilde\lambda}}\Big) +
  \frac{1}{2} \frac{d\lambda}{d{\tilde\lambda}}
  \Big\{ (D-2) \frac{d}{d\lambda} \ln A_\gamma -
  2 \frac{d}{d\lambda} \ln \alpha_\gamma\Big\}
  \nonumber \\
  &=&\frac{d\lambda}{d{\tilde\lambda}} \ \nabla_i l^i -
  \frac{d\lambda}{d{\tilde\lambda}} \frac{d}{d\lambda}
  \ln\frac{d\lambda}{d\tilde\lambda} +
  \frac{1}{2} (D-2) \frac{d\lambda}{d{\tilde\lambda}}
  \frac{d}{d\lambda} \ln A_\gamma,
  \nonumber
\end{eqnarray}  
where, in the 2nd equality, use of the expression (\ref{alpha_gamma})
for $\alpha_\gamma$ was made.
Inserting this into equation (\ref{Box_q1}),
we get

\begin{eqnarray}
  [\Box]_q f =
  \Big\{4 + 2 {\tilde\lambda} \frac{d\lambda}{d{\tilde\lambda}} \ \nabla_i l^i
  - 2 {\tilde\lambda} \frac{d\lambda}{d{\tilde\lambda}} \frac{d}{d\lambda}
  \ln\frac{d\lambda}{d\tilde\lambda} +
  {\tilde\lambda} \ (D-2) \frac{d\lambda}{d{\tilde\lambda}}
  \frac{d}{d\lambda} \ln A_\gamma\Big\} \frac{df}{d\sigma^2}.       
\end{eqnarray}  

Let us implement now condition {\bf G3}.
We require that,
if $G(S_L)$ is solution to $\Box G = 0$ in
$
\sigma^2({\tilde p}) = S_L,
$
i.e. if
$
\Box G = \big\{4 + 2 {\tilde\lambda} (\nabla_i l^i)_{|{\tilde p}}\big\}
\frac{dG}{dS_L} = 0,
$
then
${\tilde G}(\sigma^2) = G(S_L(\sigma^2))$
be solution of
$
[\Box]_q {\tilde G} = 0,
$
i.e.

\begin{eqnarray}
  4 + 2 {\tilde\lambda} \frac{d\lambda}{d{\tilde\lambda}} \ \nabla_i l^i
  - 2 {\tilde\lambda} \frac{d\lambda}{d{\tilde\lambda}} \frac{d}{d\lambda}
  \ln\frac{d\lambda}{d\tilde\lambda} +
  {\tilde\lambda} \ (D-2) \frac{d\lambda}{d{\tilde\lambda}}
  \frac{d}{d\lambda} \ln A_\gamma = 0.        
  \nonumber
\end{eqnarray}
This gives

\begin{eqnarray}\label{G3_0}
  -2 (\nabla_i l^i)_{|{\tilde p}} +
  2 \frac{d\lambda}{d{\tilde\lambda}} \ \nabla_i l^i
  - 2 \frac{d\lambda}{d{\tilde\lambda}} \frac{d}{d\lambda}
  \ln\frac{d\lambda}{d\tilde\lambda} +
  (D-2) \frac{d\lambda}{d{\tilde\lambda}}
  \frac{d}{d\lambda} \ln A_\gamma = 0.
\end{eqnarray}
Thanks to the relation
(\cite{DeWA, DeWB}; see \cite{VisA, PPV})

\begin{eqnarray}
  \nabla_a^{(p)}\big[\Delta(p, P) \nabla^a_{(p)} \sigma^2(p, P)\big] =
  2 D \ \Delta(p, P)
\end{eqnarray}
(valid for spacelike/timelike as well as null geodesics),
which gives

\begin{eqnarray}
  \nonumber
  \nabla_a \partial^a \sigma^2 =
  2 D + (\nabla_a \ln \Delta^{-1}) \ \partial^a \sigma^2
\end{eqnarray}
with $\partial^a \sigma^2 = 2 \lambda l^a$ on $\gamma$,
using (\ref{nabla_partial}) the expansion of the congruence 
can be usefully expressed in terms of the
van Vleck determinant as

\begin{eqnarray}
  \nabla_a l^a =
  \nabla_i l^i =
  \frac{D-2}{\lambda} + \frac{\partial}{\partial \lambda}\ln \Delta^{-1}
\end{eqnarray}  
(cf. \cite{VisA}).
Substituting this, equation (\ref{G3_0}) above
becomes

\begin{eqnarray}
  -2 \Bigg[\frac{D-2}{\tilde\lambda} +
    \frac{\partial}{\partial {\tilde\lambda}}\ln \Delta_S^{-1}\Bigg] +
  2 \frac{d\lambda}{d{\tilde\lambda}}
     \Bigg[\frac{D-2}{\lambda} +
    \frac{\partial}{\partial \lambda}\ln \Delta^{-1}\Bigg]
  - 2 \frac{d\lambda}{d{\tilde\lambda}} \frac{d}{d\lambda}
  \ln\frac{d\lambda}{d\tilde\lambda} +
  (D-2) \frac{d\lambda}{d{\tilde\lambda}}
  \frac{d}{d\lambda} \ln A_\gamma = 0,
  \nonumber
\end{eqnarray}  
where $\Delta_S$ is the van Vleck determinant evaluated at $\tilde p$.
This gives

\begin{eqnarray}
  \nonumber
  -2 \frac{d}{d\lambda} \ln{\tilde\lambda} -
  \frac{2}{D-2} \frac{\partial}{\partial \lambda} \ln \Delta_S^{-1} +
  2 \frac{d}{d\lambda} \ln \lambda +
  \frac{2}{D-2} \frac{\partial}{\partial \lambda} \ln \Delta^{-1} -
  \frac{2}{D-2} \frac{d}{d\lambda} \ln \frac{d\lambda}{d{\tilde\lambda}} +
  \frac{d}{d\lambda} \ln A_\gamma = 0,
\end{eqnarray}  
and then

\begin{eqnarray}
  \nonumber
  \frac{\partial}{\partial\lambda} \ln
  \Bigg[\frac{\lambda^2}{{\tilde\lambda}^2}
  \Big(\frac{\Delta_S}{\Delta}\Big)^{\frac{2}{D-2}}
  \Big(\frac{d{\tilde\lambda}}{d\lambda}\Big)^{\frac{2}{D-2}} A_\gamma\Bigg] = 0.  
\end{eqnarray}  
Thus

\begin{eqnarray}
  A_\gamma =
  C \ \frac{\tilde\lambda^2}{{\lambda}^2}
  \Big(\frac{\Delta}{\Delta_S}\Big)^{\frac{2}{D-2}}
  \Big(\frac{d{\tilde\lambda}}{d\lambda}\Big)^{-\frac{2}{D-2}},
  \nonumber
\end{eqnarray}  
where $C$ is a constant.
To determine $C$,
we note that 
in the $\lambda \rightarrow \infty$ limit,
$\tilde\lambda \simeq \lambda$,
$\Delta_S \simeq \Delta$
and
$d{\tilde\lambda}/d\lambda \rightarrow 1$;
thus,
$A_\gamma \rightarrow C$.
Since
$\alpha_\gamma = (d{\tilde\lambda}/d\lambda)^{-2} \rightarrow 1$,
the request $q_{ab} \rightarrow g_{ab}$ implies,
from the expression (\ref{qab_null}) for $q_{ab}$,
$A_\gamma \rightarrow 1$,
and then
$C = 1$.
Our expression for $A_\gamma$ is finally

\begin{eqnarray}\label{A_gamma}
    A_\gamma =
  \frac{\tilde\lambda^2}{{\lambda}^2}
  \Big(\frac{\Delta}{\Delta_S}\Big)^{\frac{2}{D-2}}
  \Big(\frac{d{\tilde\lambda}}{d\lambda}\Big)^{-\frac{2}{D-2}}.
\end{eqnarray}
In conclusion,
what we have got in this Section
is
the expression (\ref{qab_null}) for the qmetric $q_{ab}$ for null geodesics,
with the functions $\alpha_\gamma$ and $A_\gamma$ in it, defined
on the null geodesics, required to have the expressions given
by equations (\ref{alpha_gamma}) and (\ref{A_gamma}).
We notice that
no dependence on the chosen canonical
observer is present in
$\alpha_\gamma$ or $A_\gamma$.
The expression (\ref{qab_null}) for $q_{ab}$, however,
does depend on the observer, through $m^a$.

\section{$\rho$ for null geodesics (Lorentz sector)}

Using the results of previous Section,
let us proceed now to try to find out an expression for $\rho$ for
null geodesics.
In complete analogy with the timelike/spacelike case,
this quantity can be defined, in the Lorentz sector, as
(cf. equation (\ref{rho_def}))

\begin{eqnarray}\label{rho_null}
    \rho(P, l^a) =
  \bigg(\lim_{p \rightarrow P}
  \frac{[d^{D-1}V]_{q(g)}(p, P)}{[d^{D-1}V]_{q(\eta)}(p, P)}\bigg)
  _{\gamma(l^a)}.
\end{eqnarray}    
Here,
$\gamma(l^a)$ is a
null geodesic through $P$,
affinely parameterized through $\lambda = \lambda(p, P)$ with
$\lambda(P, P) = 0$,
with tangent vector $k^a = dx^a/d\lambda$ along it
which takes the value
$l^a$ at $P$, i.e. $l^a = k^a_{|P}$.
The limit is taken for $p$ approaching $P$ along $\gamma(l^a)$.
$
d^{D-1}V
$
is 
a $(D-1)$-dim volume element of a null hypersurface $\Sigma_\gamma$ through $p$,
defined by $\Phi = {\rm const}$, with $-(\partial_a \Phi)_{|p} = (k_a)_{|p}$. 
Apart from this condition on the gradient,
the hypersurface $\Sigma_\gamma$ is arbitrary.
$
[d^{D-1}V]_q
$
is the volume of that same element of hypersurface,
according to the qmetric,
with
$\Sigma_\gamma$ being null also according to the qmetric
($q^{ab} k_a k_b = 0$, as we saw before).
The index $q(g)$, or simply $q$, refers to a generic metric $g_{ab}$,
while $q(\eta)$ is for the flat case.

$d^{D-1}V$ can be written as follows
(\cite{PoiA, PadN}, e.g.).
Using the vector $m^a$ as defined in the previous Section,
we can write
the metric transverse to $k^a$ at $p$
as

\begin{eqnarray}
  \nonumber
  h_{ab} = g_{ab} + \frac{1}{2} (k_a m_b + m_a k_b).
\end{eqnarray}
Introducing the coordinates $(\lambda, \theta^A)$ for $\Sigma_\gamma$,
with the coordinates $\theta^A$ spanning the $(D-2)$-dim space
transverse to the generators of $\Sigma_\gamma$,
we have the induced metric on the $(D-2)-$dim space is given by

\begin{eqnarray}
  \nonumber
  \sigma_{AB}
  &=& g_{ab} e^a_A e^b_B
  \nonumber \\
  &=& h_{ab} e^a_A e^b_B
  \nonumber
\end{eqnarray}
in terms of the vectors
$e^a_A = \big(\frac{\partial x^a}{\partial \theta^A}\big)_\lambda$
($e^a_A$ is orthogonal to both $k^a$ and $m^a$).  
The volume element can then be written as

\begin{eqnarray}
d^{D-1}V = \sqrt{\sigma} \ d^{D-2}\theta \ d\lambda, 
\end{eqnarray}
with $\sigma = \det (\sigma_{AB})$.

Going to the qmetric,
$k^a = dx^a/d\lambda$
gets mapped to
$[k^a]_q = dx^a/d{\tilde\lambda} = (d\lambda/d{\tilde\lambda}) k^a$.
$\Sigma_\gamma$ is null also according to the qmetric,
and the metric transverse (according to $q_{ab}$) to $[k^a]_q$
is given by

\begin{eqnarray}
  [h_{ab}]_q = q_{ab} +
  \frac{1}{2} \Big([k_a]_q [m_b]_q + [m_a]_q [k_b]_q\Big),
  \nonumber
\end{eqnarray}  
with
$
[k_a]_q
= q_{ab} [k^a]_q
= \frac{1}{\alpha_\gamma} \frac{d\lambda}{d{\tilde\lambda}} k_a
= \frac{d{\tilde\lambda}}{d\lambda} k_a,
$
and
$
[m_a]_q
= \frac{d{\tilde\lambda}}{d\lambda} m_a
$
(to get
$q^{ab} [k_a]_q [m_a]_q = -2$).  
Using the expression (\ref{qab_null}) for $q_{ab}$,
we get

\begin{eqnarray}
  [h_{ab}]_q = A_\gamma h_{ab},  
\end{eqnarray}
and,
from
$
e^a_A
= \big(\frac{\partial x^a}{\partial \theta^A}\big)_\lambda
= \big(\frac{\partial x^a}{\partial \theta^A}\big)_{\tilde\lambda}
= [e^a_A]_q,
$

\begin{eqnarray}\label{sigma_q}
  [\sigma_{ab}]_q
  &=& q_{ab} [e^a_A]_q [e^b_B]_q
  \nonumber \\
  &=& [h_{ab}]_q [e^a_A]_q [e^b_B]_q
  \nonumber \\
  &=& [h_{ab}]_q e^a_A e^b_B
  \nonumber \\
  &=& A_\gamma \sigma_{ab}.
\end{eqnarray}  
The qmetric volume element
is
$
[d^{D-1}V]_q = [\sqrt{\sigma}]_q \ d^{D-2}\theta \ d{\tilde\lambda}
            = [d^{D-2} {\cal A}]_q \ d{\tilde\lambda}
$
with
$
[d^{D-2} {\cal A}]_q = [\sqrt{\sigma}]_q \ d^{D-2}\theta
$
the $(D-2)$-dim area of the element of surface
transverse to the generators according to the qmetric,
and $d^{D-2} {\cal A}$ the area according to $g_{ab}$.
By the way,
this form of $[d^{D-1}V]_q$ gives,
from
\begin{eqnarray}
  \frac{[d^{D-1}V]_{q(g)}}{[d^{D-1}V]_{q(\eta)}} =
  \frac{[d^{D-2}{\cal A}]_{q(g)}}{[d^{D-2}{\cal A}]_{q(\eta)}},
\end{eqnarray}
an equivalent manner, if one wants, to express $\rho$,
as

\begin{eqnarray}
     \rho(P, l^a) =
  \bigg(\lim_{p \rightarrow P}
  \frac{[d^{D-2}{\cal A}]_{q(g)}(p, P)}{[d^{D-2}{\cal A}]_{q(\eta)}(p, P)}\bigg)
  _{\gamma(l^a)}. 
\end{eqnarray}  
From (\ref{sigma_q}),
\begin{eqnarray}  
  [d^{D-1}V]_q
  &=& [\sqrt{\sigma}]_q \ d^{D-2}\theta \ d{\tilde\lambda}
  \nonumber \\
  &=& A_\gamma^{\frac{D-2}{2}} \sqrt{\sigma} \ d^{D-2}\theta \ d{\tilde\lambda}
  \nonumber \\
  &=& A_\gamma^{\frac{D-2}{2}} d^{D-2} {\cal A} \ d{\tilde\lambda}.
  \nonumber
\end{eqnarray}
Using, on the
the $(D-2)$-surface, orthogonal
coordinates $z^A$ such that,
chosen any one of them, $z^{\bar A}$,
it can be put in the form $z^{\bar A} = \lambda \, \chi$,
with $\chi$ such that
$\lambda \, d\chi$ is proper distance,
we can write

\begin{eqnarray}
  \nonumber
  [d^{D-1}V]_q
  = A_\gamma^{\frac{D-2}{2}} \lambda^{D-2} \ (1 + {\cal O}(\lambda^2))
  \ (d\chi)^{D-2} d{\tilde\lambda},
\end{eqnarray}
where the ${\cal O}(\lambda^2)$ term represents the effects of
curvature and is absent in flat case.
Substituting here the expression (\ref{A_gamma}) for $A_\gamma$,
we get

\begin{eqnarray}
  [d^{D-1}V]_q
  &=& {\tilde\lambda}^{D-2} \frac{\Delta}{\Delta_S}
  \ \frac{d\lambda}{d{\tilde\lambda}} \ (1 + {\cal O}(\lambda^2))
  \ (d\chi)^{D-2} d{\tilde\lambda}
  \nonumber \\
  &=& {\tilde\lambda}^{D-2} \frac{\Delta}{\Delta_S}
  \ (1 + {\cal O}(\lambda^2))
  \ (d\chi)^{D-2} d\lambda.
  \nonumber
\end{eqnarray}
Taking the limit $\lambda \rightarrow 0$ we see this quantity does
not vanish, and goes to the value

\begin{eqnarray}
  \lim_{\lambda \rightarrow 0} \ [d^{D-1}V]_q
  = L^{D-2} \frac{1}{\Delta_L(P, l^a)} \ (d\chi)^{D-2} d\lambda,
\end{eqnarray}
with
$\Delta_L(P, l^a) = \Delta({\bar p}, P)$,
where $\bar p$ is that point on the null geodesic $\gamma(l^a)$
which has $\lambda({\bar p}, P) = L$.
In the flat case,
$\Delta = 1$ identically and then $\Delta_L(P, l^a) = 1$, as we said,
and the expression above reduces to
$
\lim_{\lambda \rightarrow 0} [d^{D-1}V]_{q(\eta)} =  L^{D-2} \ (d\chi)^{D-2} d\lambda.
$
Thus,

\begin{eqnarray}\label{Delta_null}
  \rho(P, l^a)
  &=& \frac{\lim_{\lambda \rightarrow 0} [d^{D-1}V]_{q(g)}}
  {\lim_{\lambda \rightarrow 0} [d^{D-1}V]_{q(\eta)}}
  \nonumber \\
  &=& \frac{1}{\Delta_L(P, l^a)}.
\end{eqnarray}
We obtain then, in the null case, that same form we found in the
timelike/spacelike case.
Since $l^a$ is assigned with the null geodesic at start,
we notice that,
even if the qmetric $q_{ab}$ does depend on the chosen observer (through $m^a$),
no dependence on the observer is left in $\rho$.

For timelike/spacelike geodesics,
we gave an expansion of $\Delta(p, P)$ in powers
of $l = \sqrt{\epsilon \sigma^2}$ (equation (\ref{expansion})).
For (affinely parameterized) null geodesics,
$\Delta(p, P)$ can be analogously expanded in powers
of $\lambda$ as (\cite{DeWA}; \cite{Xen, VisA, PPV})

\begin{eqnarray}
  \Delta(p, P) = 1 + \frac{1}{6} \lambda^2 R_{ab}(P) l^a l^b
                 + o(\lambda^2 R_{ab}(P) l^a l^b). 
\end{eqnarray}
For $l^a$ in a neighbourhood of $0$,
this definitely gives

\begin{eqnarray}
  \Delta_L(P, l^a) = 1 + \frac{1}{6} L^2 R_{ab}(P) l^a l^b
                     + o(L^2 R_{ab}(P) l^a l^b),
\end{eqnarray}  
and

\begin{eqnarray}
  \rho(P, l^a)
  = 1 - \frac{1}{6} L^2 R_{ab}(P) l^a l^b + o(L^2 R_{ab}(P) l^a l^b).
\end{eqnarray}  
This expression for $\rho$ is analogous
to that reported above for timelike/spacelike geodesics
(equation (\ref{rho_expansion})),
and coincides with the expression which has been found
through recourse to Euclidean sector
\cite{Pad04, Pad06, Pad08, Pad10}.

\section{Conclusions}

The paper has tried to face the problem of having
a working definition of the effective metric $q_{ab}$
in the case of null geodesics.
A notion of $q_{ab}$ for null geodesics has been introduced
and an expression for it has been provided (equation (\ref{qab_null})).
This, adding to the existing expressions for $q_{ab}$ for timelike
and spacelike geodesics \cite{StaA},
is supposed to complete the description of $q_{ab}$ in
spacetime.
Using of these formulas, 
an expression for $\rho$ for the null case is derived
remaining entirely in the Lorentz sector,
i.e. without making use of Euclideanization.
What we obtain coincides
with the formula
derived
\cite{Pad04, Pad06, Pad08, Pad10}
with the latter.
The formula for $\rho$ for null case,
joined with the formulas for timelike/spacelike cases,
provides in principle a complete account of $\rho$
in spacetime.

{\it Acknowledgements.}
I wish to thank Sumanta Chakraborty and Dawood Kothawala
for careful reading of the manuscript
and for their comments.



\begin{thebibliography}{00}

\bibitem{KotE}
D. Kothawala,
``Minimal length and small scale structure of spacetime'',
Phys. Rev. D {\bf 88} (2013) 104029,
arXiv:1307.5618.

\bibitem{KotF}
D. Kothawala, T. Padmanabhan,
``Entropy density of spacetime as a relic from quantum gravity'',
Phys. Rev. D {\bf 90} (2014) 124060,
arXiv:1405.4967.

\bibitem{StaA}
D.J. Stargen, D. Kothawala,
``Small scale structure of spacetime: van Vleck determinant and equi-geodesic 
surfaces'',
Phys. Rev. D {\bf 92} (2015) 024046,
arXiv:1503.03793.

\bibitem{Pad04}
T. Padmanabhan,
``Distribution function of the atoms of spacetime and the nature of gravity'',
Entropy {\bf 17} (2015) 7420,
arXiv:1508.06286.

\bibitem{Pad06}
T. Padmanabhan,
``Exploring the nature of gravity'',
arXiv:1602.01474 (2016).

\bibitem{Pad08}
T. Padmanabhan,
``The atoms of space, gravity and the cosmological constant'',
Int. J. Mod. Phys. D {\bf 25} (2016) 1630020, 
arXiv:1603.08658.

\bibitem{Pad10}
T. Padmanabhan,
``The atoms of spacetime and the cosmological constant'',
Journal of Physics: Conf. Series {\bf 880} (2017) 012008,
arXiv:1702.06136.

\bibitem{PesK}
A. Pesci,
``Spacetime atoms and extrinsic curvature of equi-geodesic surfaces'',
arXiv:1511.08665 (2015).

\bibitem{Syn}
J.L. Synge,
{\it Relativity: The general theory}
(North-Holland, Amsterdam, 1960).

\bibitem{vVl} 	
J.H. van Vleck,
``The correspondence principle in the statistical interpretation of
quantum mechanics'',
Proc. Nat. Acad. Sci. USA {\bf 14} (1928) 178.

\bibitem{Mor}
C. Morette,
``On the definition and approximation of Feynman's path integrals'',
Phys. Rev. {\bf 81} (1951) 848.

\bibitem{DeWA}
B.S. DeWitt, R.W. Brehme,
``Radiation damping in a gravitational field'',
Annals Phys. {\bf 9} (1960) 220.

\bibitem{DeWB}
B.S. DeWitt,
{\it The dynamical theory of groups and fields}
(Gordon and Breach, New York, 1965).

\bibitem{Xen}
S.M. Christensen,
``Vacuum expectation value of the stress tensor in an arbitrary curved 
background: The covariant point-separation method'',
Phys. Rev. D {\bf 14} (1976) 2490.

\bibitem{VisA}
M. Visser,
``van Vleck determinants: geodesic focussing and defocussing in Lorentzian
spacetimes'',
Phys. Rev. D {\bf 47} (1993) 2395,
hep-th/9303020.

\bibitem{PPV}
E. Poisson, A. Pound, I. Vega,
``The motion of point particles in curved spacetime'',
Liv. Rev. Rel. {\bf 14} (2011) 7,
arXiv:1102.0529.

\bibitem{KotG}
D. Kothawala,
``Intrinsic and extrinsic curvatures in Finsler{\it esque} spaces'',
Gen. Rel. Grav. {\bf 46} (2014) 1836,
arXiv:1406.2672.

\bibitem{PoiA}
E. Poisson,
{\it A relativist's toolkit}
(Cambridge Univ. Pr., Cambridge UK, 2004).

\bibitem{PadN}
T. Padmanabhan,
{\it Gravitation: Foundations and frontiers}
(Cambridge Univ. Pr., Cambridge UK, 2010).


\end{thebibliography}
\end{document}